# A Dynamic Model for Frequency Response Optimization in Photovoltaic Visible Light Communication

Shuyan Chen, *Graduate Student Member, IEEE,* Hui Yu, Ni Zhao, and Lian-Kuan Chen, *Senior Member, IEEE*

*Abstract*— Photovoltaic (PV) modules are recently employed in photovoltaic visible light communication (PVLC) for simultaneous energy harvesting and visible light communication. A PV-based receiver features large signal output, easy optical alignment, and self-powered operation. However, PV modules usually have a severe bandwidth limitation when used as passive photodetectors. Hence, a fundamental understanding of a PV module's frequency response characteristics is essential to achieve a higher communication performance without applying external power. In this paper, we systematically investigate the internal impedance dynamic of PV modules and how that affects their frequency response characteristics under different illuminances. We propose a simplified yet accurate dynamic PV mode AC detection model to capture the frequency response characteristics of a PVLC receiver. The model is validated with the impedance spectroscopy characterization methodologies. Experimental results show that a PV module's internal resistance and capacitance depend on incident illuminance, affecting PV's frequency response. The bandwidth is exacerbated under indoor environments with low illuminance levels due to the increment of internal resistance for PV modules. The RC constant can be reduced for PVLC receivers working near open-circuit voltage conditions by adding a moderate local light to decrease the internal resistance value. For practical implementation, PVLC receivers will employ a load for data recovery. We show that adjusting the forward bias conditions can simultaneously reduce the resistance and capacitance values. With the optimization of equivalent trans-impedance, the data rate of a Cadmium telluride (CdTe) PV module achieves a 3.8 times enhancement under 200 lux. We also demonstrate that the BER of a 5-Mbit/s eight-level pulse amplitude modulation (PAM8) signal can be reduced from $9.8\times10^{-2}$ to $1.4\times10^{-3}$ by maximizing the transimpedance gain-bandwidth product.

## NOMENCLATURE

| | |
|---|---|
| $I_{PH}$ | Photocurrent |
| $i_{PH}(t)$ | Time-domain photocurrent signal |
| $I_D$ | Diode forward current |
| $I_0$ | Diode reverse saturation current |
| $V_{DC}$ | Bias voltage across the terminals of the PV module |
| $v(t)$ | Time-domain photovoltage signal |
| $V_T$ | Equivalent thermal voltage |
| $r$ | Diode small signal model |
| $R_{SH}$ | Shunt resistance |
| $R_S$ | Series resistance |
| $R_L$ | Load resistance |
| $R_P$ | Equivalent parallel resistance |
| $R_{rec}$ | Recombination resistance |
| $C_P$ | Equivalent parallel capacitance |
| $C_d$ | Diffusion capacitance |
| $C_{dl}$ | Depletion capacitance |
| $Z$ | PV module internal impedance |
| $Z_{TI}$ | PVLC receiver impedance |
| $n$ | Diode ideality factor |
| $q$ | Electron charge |
| $k_B$ | Boltzmann's constant |
| $T$ | Temperature in kelvin |
| $j$ | Imaginary unit |
| $\omega$ | Angular frequency |
| $\mathscr{F}(\cdot)$ | Fourier transform |
| $H(\cdot)$ | Transimpedance transfer function |
| $Re\{\cdot\}$ | Real part of a complex number. |
| $Im\{\cdot\}$ | Imaginary part of a complex number. |

## I. INTRODUCTION

PHOTOVOLTAIC visible light communication (PVLC) is a promising technology that employs photovoltaic (PV) modules as a photodetector in visible light communication (VLC) receivers. The communication capability of PV modules using unbiased signal detection enables simultaneous energy harvesting and communication function [1]. Compared with photodiodes (PD) commonly used for signal detection in VLC, PV modules significantly alleviate the alignment requirement due to their large physical area. The PD receiver has a very small physical area, thus restricting its detection efficiency for diffused light. With the booming utilization of renewable power, photovoltaic (PV) modules are widely deployed for energy harvesting in many electronic devices. Extending PV's use to VLC applications attracts new research interest [2]-[4].

For PVLC, the PV module's biasing conditions will influence its performance as a photodetector system [5]. The solar panel can work in photoconductive mode (PC) and photovoltaic mode (PV). The former and the latter correspond to the cases with and

Manuscript received xxx xx, 2023; revised xxx xx, 2023; accepted xxx xx, 2023. Date of publication xxx xx, 2023; date of current version xxx xx, 2023. This work was supported in part by the HKSAR RGC grant (GRF 14207220). (Corresponding author: Lian-Kuan Chen.)

Shuyan Chen, and Lian-Kuan Chen are with the Department of Information Engineering, The Chinese University of Hong Kong, Shatin, N. T., Hong Kong SAR. (e-mail: lkchen@ie.cuhk.edu.hk).

Hui Yu, and Ni Zhao are with the Department of Electronic Engineering, The Chinese University of Hong Kong, Shatin, N. T., Hong Kong SAR.

Color versions of one or more figures in this letter are available at https://doi.org/xx. xxxx/JLT. xxxx. xxxxxxxx.

Digital Object Identifier xx. xxxxx/JLT. xxxx. xxxxxxxx.

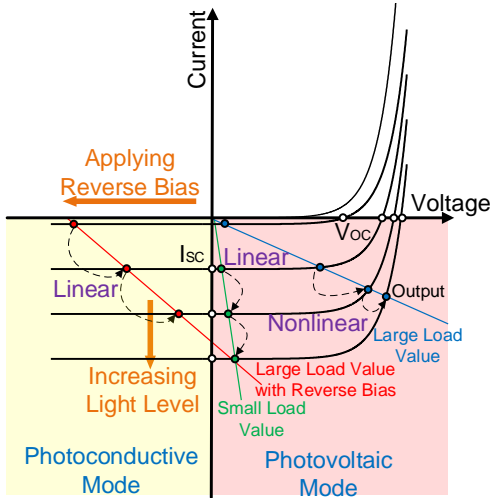

Fig. 1. The photovoltaic O/E detection principle.

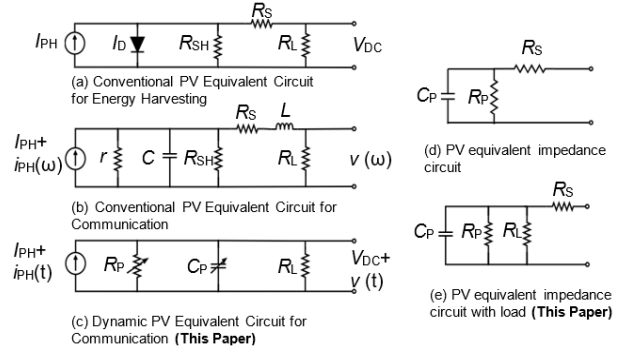

Fig. 2. A review chart on the development of photovoltaic AC model.

without reverse bias, respectively. In PC mode, the reverse bias provides higher sensitivity, wider bandwidth, and improved linearity, as shown by the red line in Fig. 1. Conversely, no bias is applied in PV mode. Thus, it self-generates forward bias voltage. The solar panel often works in PV mode for energy harvesting; therefore, it is usually regarded as a direct-current (DC) component, and its DC characteristic is well-studied, while alternate-current (AC) characteristics are much less investigated [6]–[8].

One challenging issue in PVLC is the severe band-limited effect of PV modules, limiting the overall performance. Conventionally, reverse biases are applied to PV modules allowing them to work in the PC mode [9][10]. However, PC mode operation usually requires a very large power consumption that causes hotspots in PV cells and may damage the PV modules. Another approach to increase the bandwidth is to use a transimpedance amplifier (TIA) to convert photocurrent (the green line in Fig. 1) into voltage signals at the cost of energy consumption [12]. However, PV modules usually have different DC characteristics, thus necessitating individual TIA parameters optimization for each module, resulting in higher costs. It is also challenging for the TIA output signal to stay in the linear region under a wide range of illuminance. Therefore, a PV module that works in the self-powered PV mode with larger bandwidth is highly desirable. Optimizing the bandwidth enhancement for self-powered PVLC receivers requires a comprehensive understanding of the frequency response characteristics of PV modules with direct voltage output signals (the blue line in Fig. 1).

In this paper, we systematically investigated the frequency response characteristics by analyzing the dynamic internal resistance and capacitance. We show that the resistance and capacitance values vary widely with the illuminance and forward voltage variation, affecting the PV's 3-dB bandwidth. A new PV-mode impedance dynamic model is proposed to characterize the dominant factors that affect the frequency response of a PVLC receiver under different illuminance. Experimental results show that indoor illuminance conditions (<1000 lux) guarantee low noise and high AC conversion efficiency. However, the performance is mainly limited by the exacerbated 3-dB bandwidth due to the increased internal resistance for RC constant-limited PVLC receivers. Based on the newly developed model, we propose a novel bias-voltage-adaption (BVA) method that can simultaneously reduce the resistance and capacitance by applying an optimized load value smaller than the PV internal resistance. With BVA, a Cadmium telluride (CdTe) PV module's data rate improved from 1.4 Mbit/sec to 5.2 Mbit/sec under 200 lux in self-powered PV mode. We also demonstrate that the bit-error-ratio (BER) of a 5-Mbit/s eight-level pulse amplitude modulation (PAM8) signal can be reduced from $9.8\times10^{-2}$ to $1.4\times10^{-3}$ by maximizing the transimpedance gain-bandwidth product.

The remainder of the paper is structured into five sections: Section II reviews the current PVLC AC detection model and proposes a novel photovoltaic AC dynamic model. The experimental setup we used for the model validation is introduced in Section III. Section IV discusses the experimental results. The PVLC performance is evaluated in terms of bandwidth, data rate, and BER under different illuminances. Finally, Section V gives the conclusion of this work.

## II. PRINCIPLES OF A DYNAMIC PHOTOVOLTAIC AC MODEL

In this section, we introduce the AC detection model for PV modules. Firstly, we review the conventional AC photovoltaic models and state the motivations for developing a new PV-mode impedance dynamic model. Secondly, the new model that can capture dynamic frequency response characteristics is derived and introduced. Finally, the VBA principle for bandwidth optimization is discussed. The notations commonly used in this paper are defined in the nomenclature.

### A. Conventional PV Model for PVLC

The first PV communication model originates from the conventional single-diode model used for solar cell modeling (Fig.2(a)) [13]. It contains a current source ($I_{PH}$) and a diode for modeling the forward current ($I_D$). For practical solar cells, it also includes a shunt resistance ($R_{SH}$) and serial resistance ($R_S$) for modeling the output voltage and current loss. The diode component is linearized into an equivalent small-signal resistance for communication modeling. A capacitor is also added in parallel to the current source to model the band-limited effect of the PV module, as shown in Fig.2(b) [14]. Finally, an inductor is added in series to the load $R_L$ to model the inductance of the wires attached to the PV module's electrodes.

The components in the equivalent circuit model convert the generated photocurrent signal into voltage signals. As shown in Fig.2, The DC and AC components generated by the current source are denoted as $I_{PH} + i_{PH}(\omega)$. While the output is the DC and AC voltage on the load, which is $V_{DC} + v(\omega)$. Thus the frequency response is written as:

$$H(\omega) = \left|\frac{v(\omega)}{i_{PH}(\omega)}\right|^2 = \left|\frac{\frac{R_L}{R_X}}{\frac{1}{r} + j\omega C + \frac{1}{R_{SH}} + \frac{1}{R_X}}\right|^2, \quad (1)$$

where $R_X$ is denoted as:
$$R_X = R_S + R_L + j\omega L. \quad (2)$$

Eq. (1) is the conventional model that clearly illustrates the AC detection characteristics for PV modules. The model, however, does not reveal frequency response characteristics under different illuminance conditions. This is because the model assumes that $r$ and $C$ are constants, which is valid under reverse bias conditions. For PV mode, we show that $r$ and $C$ also depend on $I_{PH}$ and $V_{DC}$. Another issue is that the dominant factors affecting frequency response are less discussed. In the next part, we develop an accurate yet simple communication model for the PV module and investigate the PV module's dynamic frequency response characteristics.

*B. PV-mode Impedance Dynamic Model*

To facilitate the estimation of frequency response for PV modules under different illuminance, we propose a new dynamic model, as shown in Fig.2(c). It contains a current source and three dominant components parallel to the current source: $R_P$, $C_P$, and $R_L$. The current source models the PV module detection of DC and AC optical signals continuously in the time domain, and the output signal is $I_{PH}+i(t)$. $R_P$ includes the parallel resistance components, $r$ and $R_{SH}$ in the conventional PVLC model for measurement simplicity. In this new frequency response model, we neglected the effect of $R_S$ and $L$. Since the series resistance $R_S$ is usually more than a thousand times smaller than the shunt resistance $R_{SH}$ [15]-[17], we can assume its effect on frequency response is minor and neglect it. The inductor in the $R_X$ of Eq. (1) can also be neglected within a reasonable input frequency range [18]. For PVLC receivers, a part of $I_{PH}+i(t)$ will flow through $R_L$ and generate $V_{DC}+v(t)$. Therefore, the transimpedance transfer function of the new PV model shown in Fig.2(c) is given by the ratio of $\mathcal{F}(v(t))$ and $\mathcal{F}(i_{PH}(t))$:

$$H(\omega) = \left|\frac{\mathcal{F}(v(t))}{\mathcal{F}(i_{PH}(t))}\right|^2 = \left|\frac{R_L}{\frac{R_L}{R_P} + R_L * j\omega C_P + 1}\right|^2. \quad (3)$$

Next, we characterize the dynamics of Eq. (3) due to the variation of $V_{DC}$. The combined parallel resistance value can be written as:

$$R_P = \frac{rR_{sh}}{R_{sh} + r}. \quad (4)$$

When PV modules operate in forward-bias conditions, the diode component is linearized as a small-signal resistance model $r$, which is denoted as:

$$r = \frac{nV_T}{I_D}, \quad (5)$$

where the diode current $I_D$ is given by:
$$I_D = I_0\left(e^{\frac{qV_{DC}}{nk_BT}} - 1\right). \quad (6)$$

Eq. (4), Eq. (5), and Eq. (6) indicate that a PV module's parallel resistance is dynamic because $R_P$ is inversely proportional to $I_D$. Consequently, the transfer function depends both on the input signal frequency and $V_{DC}$, which can be written as:

$$H(\omega, V_{DC}) = \left|\frac{1}{\frac{1}{R_{SH}} + \frac{nV_T}{I_0\left(e^{\frac{qV_{DC}}{nk_BT}} - 1\right)} + j\omega C_P + \frac{1}{R_L}}\right|^2. \quad (7)$$

Eq. (7) indicates that the frequency response characteristics (transimpedance gain, 3-dB bandwidth, etc.) of a PVLC receiver are susceptible to incident illuminance. For instance, when working under open-circuit voltage conditions, the bandwidth of a PVLC receiver increases due to the decrement of $R_P$ (see experimental results in section IV-B). This trend also applies to PV-mode-based modulators, which recently emerged for low-power sensing [19]. It is also worth noting that $C_P$ also changes with $V_{DC}$. The effect of $C_p$ on the frequency response will be discussed in the next section.

*C. Internal Impedance Characteristics of PV Modules*

In order to accurately extract the parameter values of $R_P$ and $C_P$, the impedance characteristics under different $V_{DC}$ conditions must be investigated. For solid-state PV cells, $R_P$ mainly represents the recombination process resistance ($R_{rec}$). $C_P$ represents the diffusion capacitance ($C_d$) in forward bias or the depletion capacitance ($C_{dl}$) in reverse bias [20]. Therefore, in PV mode operations, $R_P$ can be denoted as:

$$R_P = R_{rec} = \frac{\tau_n}{C_d}, \quad (8)$$

where $\tau_n$ is the minority carrier lifetime [2]. The parallel capacitance in forward bias is given by [20]:

$$C_P = C_d = A\frac{q^2Ln_0}{k_BT}\exp\left(\frac{qV_{DC}}{k_BT}\right), \quad (9)$$

where $L$ is the thickness of the junction layer, $n_0$ is the minority carrier density in equilibrium, and $A$ is the physical area of the solar cell. Eq. (8) and Eq. (9) show that the capacitance is proportional to $V_{DC}$ while the resistance is inversely proportional to $V_{DC}$. Therefore, the frequency response of a PVLC receiver depends on the illuminance and the biasing conditions.

Fig. 2(d) shows the circuit diagram formed for parameter estimation [20]. The parameter values of $R_P$, $C_P$, and $R_S$ are extracted using an impedance analyzer to measure the impedance spectra and fit the impedance spectra with the PV module's internal impedance. $Z$ is given by:

$$Z = R_S + \frac{R_P}{1 + j\omega C_P R_P}. \quad (10)$$

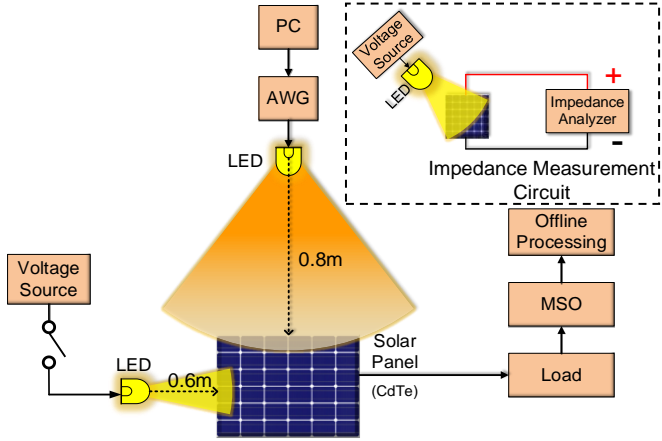

Fig. 3. Experimental setup for solar panel VLC system. AWG: arbitrary waveform generator, MSO: mixed-signal oscilloscope, Inset: Impedance measurement system.

Eq. (10) shows that series resistance $R_S$ will not affect the spectral features in the impedance spectra but only causes a horizontal displacement of the spectra. When a load is attached to the PV module to extract the received signal, the total impedance changes, as shown in Fig. 2(e). The PVLC receiver impedance is then given by

$$Z_{TI} = R_S + \frac{R_P \| R_L}{1 + j\omega C_P (R_P \| R_L)}. \quad (11)$$

where $R_P \| R_L$ is the resistance value of $R_P$ and $R_L$ in parallel. Eq. (11) shows that the RC constant and transimpedance of a PVLC receiver depend strongly on $V_{DC}$. The transimpedance corresponds to the gain of the PV module, converting current to an output voltage. Changing the $R_L$, $I_{PH}$, or biasing condition (via applying an external bias to the PV module) will change the received signal characteristics in PVLC.

### III. EXPERIMENTAL SETUP

Fig. 3 illustrates the experimental setup of the PV-module-based VLC systems. On the transmitter side, AC waveforms are generated by an arbitrary waveform generator (AWG, Siglent SDG 6022X) with a DC bias superimposed internally. The output electrical signals drive a light-emitting diode (LED, OSRAM LUW W5AM) to generate visible light signals. The measured 3-dB bandwidth of LED is 1.2 MHz. A PV module is connected to a receiving circuit at the receiver side. A resistor load is used in the receiving circuit to extract the converted electrical power from the PV module. After 0.8-m free-space transmission, the optical signal is detected by the PV module (Cadmium telluride, size 165×133 mm$^2$). A mixed-signal oscilloscope (MSO, Tektronix MSO 4054) records the voltage on the load for subsequent offline signal processing—frame synchronization, Least Mean Square (LMS) linear equalization, and PAM8 demodulation. The wires used at the receiver side are all warped with aluminum foil to shield the ambient electromagnetic waves for more accurate frequency response measurement. A local LED (OSRAM LUW W5AM) driven by a DC voltage is placed beside the PV module at a distance of 0.6 m to adjust the operation illuminance of the PV module. A lux meter (Smart sensor, AS823) is used to measure the illuminance. An impedance analyzer (BioLogic, SP200) measures the internal parameters of the PV module. We have chosen three typical illuminances for investigation: 200 lux, 600 lux, and 1000 lux corresponding to home scenarios, supermarket scenarios, and mechanical workshop scenarios, respectively. For performance evaluation, we estimated the BER performance with a forward error correction (FEC) threshold value of 3.8×10$^{-3}$.

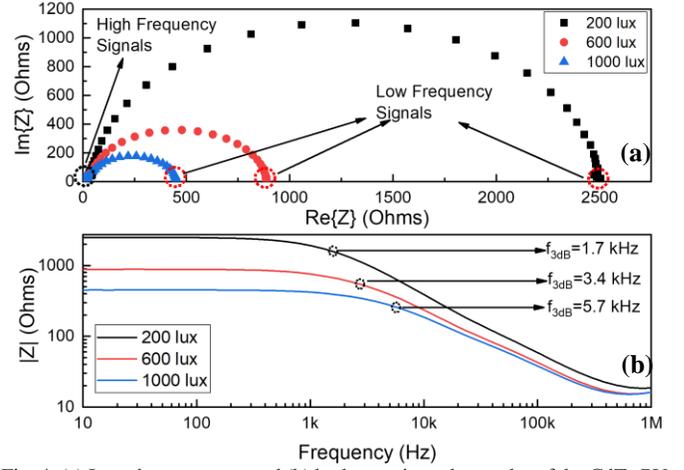

Fig. 4. (a) Impedance spectra and (b) bode transimpedance plot of the CdTe PV module under different illuminance.

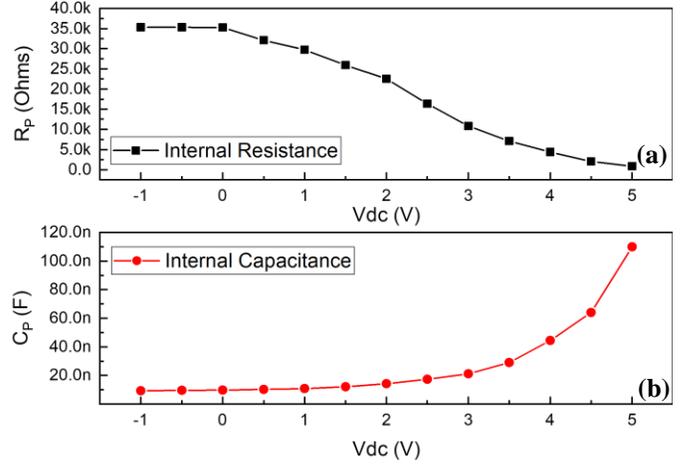

Fig. 5. (a) Internal resistance and (b) capacitance of the CdTe PV module with respect to variation of $V_{DC}$ under dark condition (0 lux).

### IV. EXPERIMENTAL DEMONSTRATION

In this section, we investigate the proposed model's efficiency, the PV modules' frequency response, and communication performance in PVLC. Firstly, we estimate the impedance spectra and internal parameter variation of CdTe PV modules under different illuminance and bias. Secondly, the frequency responses of PV modules with different $Z_{TI}$ and under different illuminances are investigated. Finally, the communication performance with optimized $Z_{TI}$ is demonstrated.

#### A. Impedance spectra and internal parameters

Fig. 4(a) shows the open-circuit impedance spectra of the CdTe PV module working under different illuminances. The values of $R_S$ can be estimated in the impedance spectra by reading the $Re\{Z\}$ when $Im\{Z\}$ is zero at high-frequency input signals [20]. Similarly, $R_P$ is estimated when $Im\{Z\}$ is zero at low-frequency inputs. The $R_S$ is measured to be around 10 Ω for all the illuminance because series resistance is mainly caused by metal contacts and thus is relatively stable. $R_P$ decreases from 2.5 kΩ to 0.41 kΩ when increasing the illuminance from 200 to 1000 lux. The results agree with the assumption of the proposed simplified model, and the increment of $R_P$ agrees well with the definition of internal resistance for PV modules. Next, the impedance spectra are converted into a bode plot by calculating the modulus of $Z$ to estimate the 3-dB bandwidth of the PV module. As shown in Fig. 4(b), the 3-dB bandwidth increases from 1.7 kHz to 5.7 kHz when increasing the illuminance from 200 lux to 1000 lux. However, the transimpedance gain decreases from 2500 to 410, which shows a tradeoff between the 3-dB bandwidth and the signal gain, usually noted as the Gain-Bandwidth tradeoff in photodetectors [22].

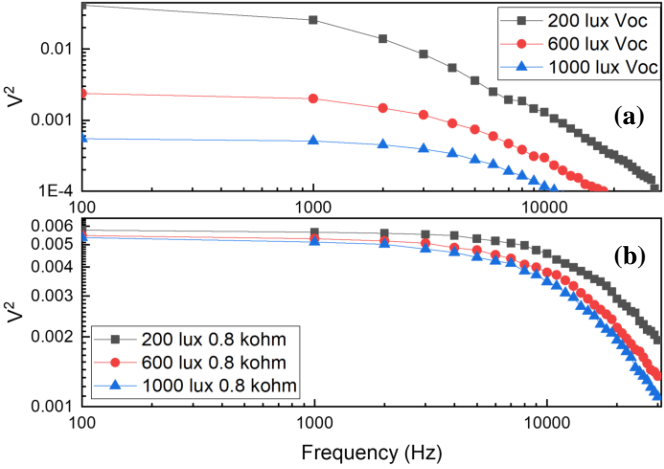

Fig. 6. Frequency response of the CdTe PV module working at (a) open circuit and (b) $R_L$=0.8kΩ, under different illuminance.

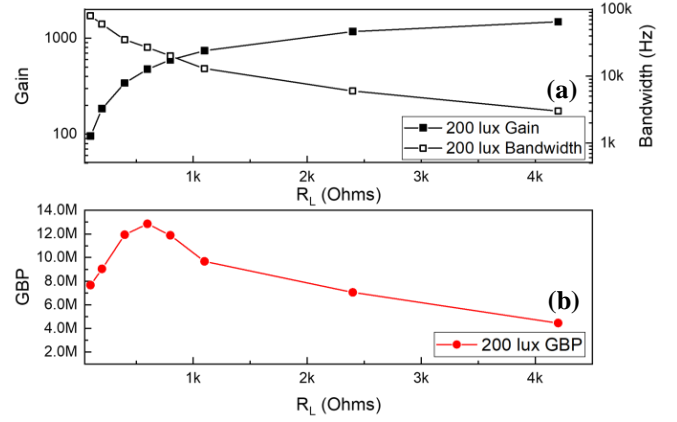

Fig. 7. (a) Gain, bandwidth and (b) Gain-Bandwidth product variation of the CdTe PV module with different $R_L$ used.

The internal parameters with respect to $V_{DC}$ are estimated by measuring the frequency spectra of the PV module under different bias voltages. Fig. 5 shows the measured internal resistance and capacitance of the CdTe PV module. It clearly illustrates that $R_P$ and $C_P$ are stable under reverse bias conditions but vary with respect to $V_{DC}$. When increasing the $V_{DC}$ from -1 V to 5 V, $R_P$ decreases sharply from 35 kΩ to 500 Ω, while $C_P$ increases from 10 nF to 90 nF. The increment of $C_P$ is much smaller when compared with the decrease in $R_P$. Thus, the overall RC constant decreases, and the 3-dB bandwidth increases, as shown in Fig. 4(b).

### B. Frequency response of PV modules

We have shown in the previous section that the open-circuit voltage bandwidth of PV modules increases with respect to the increment of illuminance due to the decrement of internal resistance. However, the measured bandwidth increment is limited. We found that the reason is that as diffusion capacitance increases with the increment of illuminance, as shown in Fig. 5, the overall change in RC constant is less under different illuminance, thus less change in bandwidth. It would be desirable if both $R_P$ and $C_P$ could be reduced simultaneously, enabling significant bandwidth enhancement for PVLC receivers. In this paper, we proposed the bias-voltage-adaption (VBA) scheme by applying a variable $R_L$ smaller than $R_P$ to decrease $V_{DC}$ and $C_P$ correspondingly, resulting in a simultaneous reduction in overall $R$ and $C$. For PVLC receivers working under open circuit conditions, $V_{DC}$ is mainly generated by the current flowing through $R_P$. When a variable load is added to extract voltage signals, $V_{DC}$ is generated by the current flowing through $R_P||R_L$. The $V_{DC}$ will be smaller than the $V_{DC}$ under open-circuit conditions because $R_P||R_L$ is smaller than $R_P$ as the photocurrent is the same under the same illuminance. Consequently, as shown in Fig. 5, $C_P$ decreases. Although $R_P$ increases with reduced $V_{DC}$, the overall $R$ is still smaller than the open-circuit $R_P$ because $R_L$ is smaller than open-circuit $R_P$. Thus, the overall $R$ and $C$ reduce, leading to significant bandwidth enhancement. Fig. 6 shows the experimental verification of PV module frequency response under open-circuit conditions and with the VBA scheme. Under open-circuit conditions, the 3-dB bandwidth and signal amplitude variation agree well with the measured impedance characteristics, as shown in Fig. 6(a). The 3-dB bandwidth under 200 lux is 1.7 kHz, the same as that derived from impedance analysis (Fig. 4(b)). By applying an $R_L$=800 Ω, smaller than $R_P$ (2.5 kΩ), the bandwidth under 200 lux increases significantly from 1.7 kHz to 25 kHz. However, the increased illuminance will cause bandwidth degradation rather than bandwidth enhancement for a fixed RL smaller than RP, as in the open-circuit case. The bandwidth drops because $C_P$ increases, but the decrement of $R_P$ will not affect $R_P||R_L$ much as the fixed $R_L$ is smaller than $R_P$. The overall RC constant will increase with the increased illuminance causing bandwidth degradation. Thus the $R_L$ should be optimized for various illuminances. As shown in Fig. 6(b), the 3-dB bandwidth of the PV module decreases from 25 kHz to 18 kHz with the increment of illuminance.

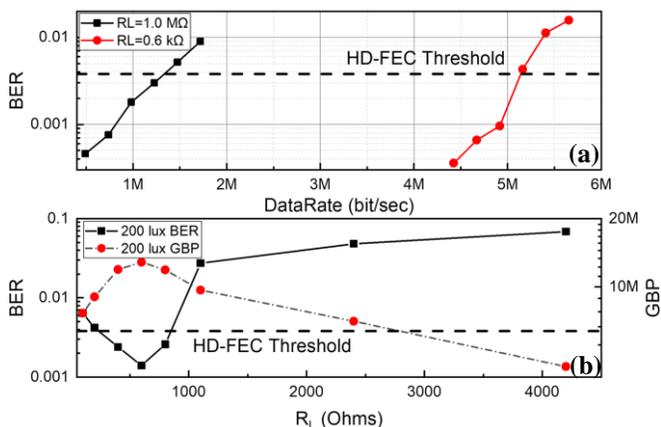

Fig. 8. (a) Datarate of the CdTe PV module working with different $R_L$ and (b) BER of a 5 Mbit/s signal when detecting with different $R_L$. Modulation Index=0.43

*C. Communication performance*

The previous section shows that 3-dB bandwidth can be increased by optimizing $R_L$. However, varying $R_L$ will also change the transimpedance (gain) value, as depicted in Eq. (11). Therefore, we measured the gain and bandwidth to find an $R_L$ that maximizes the gain-bandwidth product. Fig. 7(a) shows the gain-bandwidth tradeoff when tuning $R_L$ from 100 Ω to 4.2 kΩ. Although the bandwidth can potentially increase to 80 kHz using a 100-Ω resistor, decreasing the load will severely degrade the signal amplitude due to the gain drop, a tradeoff between gain and bandwidth. Fig. 7(b) shows the relationship between GBP and $R_L$, showing that the GBP maximizes around 600 Ω.

We then investigate the communication performance with the optimized $R_L$ by evaluating the BER of the received data. We first evaluate the data rate enhancement as shown in Fig. 8(a). The achievable data rate increased from 1.4 Mbit/s to 5.2 Mbit/s due to the significant increase in PV bandwidth. The BER after 5 Mbit/s increases more because the baud rate exceeds the bandwidth of the LED. Albeit indoor scenarios using LED lighting are considered in this study, a higher data rate can be achieved using lasers or micro-LEDs as in optical wireless communication [23][24]. Fig. 8(b) shows the BER variation when using different $R_L$ for a 5-Mbit/s PAM8 signal. The BER variation further attests to the GBP variation when using $R_L$ with different values. With the increase of $R_L$ from 100 Ω to 600 Ω, the signal SNR improves, and BER decreases. However, further increasing the $R_L$ will cause more severe inter-symbol interference to the received pulses, resulting in BER degradation and detection failure.

V. CONCLUSIONS

In this paper, we provide a systematical and comprehensive investigation of the frequency response characteristics of the PV module by analyzing the internal resistance and capacitance. We show that the resistance and capacitance value varies widely with the illuminance and forward voltage variation, thus affecting the 3-dB bandwidth. Experimental results show that a PV module's bandwidth is severely limited under indoor illuminance levels due to the limitation of internal resistance for RC constant-limited PV modules. We propose a bias-control scheme by using variable load for PVLC receiver optimization, the resistance and capacitance value can be simultaneously reduced. With the optimized RC constant, a CdTe PV module's data rate improved from 1.4 Mbit/sec to 5.2 Mbit/s under 200 lux. We also demonstrate that the bit-error-ratio (BER) of a 5-Mbit/s eight-level pulse amplitude modulation (PAM8) signal reduces from $9.8\times10^{-2}$ to $1.4\times10^{-3}$, a nearly two-order magnitude reduction, by maximizing the transimpedance gain-bandwidth product.